\documentclass[thmsb,11pt]{article}%
\usepackage{amsmath}
\usepackage{graphicx}
\usepackage{amsfonts}
\usepackage{amssymb}%
\newtheorem{theorem}{Theorem}
\newtheorem{acknowledgement}[theorem]{Acknowledgement}

\newtheorem{corollary}[theorem]{Corollary}

\newtheorem{proposition}[theorem]{Proposition}
\newtheorem{remark}[theorem]{Remark}

\newenvironment{proof}[1][Proof]{\noindent\textbf{#1.} }{\ \rule{0.5em}{0.5em}}
\begin{document}

\title{Symplectic Non-Squeezing Theorems, Quantization of Integrable Systems, and
Quantum Uncertainty}
\author{Maurice A. de Gosson\medskip\\Universit\"{a}t Potsdam, Inst. f. Mathematik \\Am Neuen Palais 10, D-14415 Potsdam\\and\\Universidade de S\~{a}o Paulo, IME\\CEP 05508-900 \ S\~{a}o Paulo\\E-mail address: maurice.degosson@gmail.com}
\maketitle

\begin{abstract}
The ground energy level of an oscillator cannot be zero because of
Heisenberg's uncertainty principle. We use methods from symplectic topology
(Gromov's non-squeezing theorem, and the existence of symplectic capacities)
to analyze and extend this heuristic observation to Liouville-integrable
systems, and to propose a topological quantization scheme for such systems,
thus extending previous results of ours.

\end{abstract}

\section{Introduction}

The fact that the ground energy level of a harmonic oscillator is different
from zero is heuristically justified in the physical literature by the
following observation: since Heisenberg's uncertainty relation $\Delta p\Delta
x\geq\frac{1}{2}\hbar$ prevent us from assigning simultaneously a precise
value to both position and momentum, the oscillator cannot be at rest. To show
that the lowest energy has the value$\frac{1}{2}\hbar\omega$ predicted by
quantum mechanics one then argues as follows: since we cannot distinguish the
origin from a phase plane trajectory whose all points lie inside the double
hyperbola $px<\frac{1}{2}\hbar$ , we must require that at least one point
$(x,p)$ of that trajectory is such that $|px|\geq\frac{1}{2}\hbar$;
multiplying both sides of the trivial inequality
\[
\frac{p^{2}}{m\omega}+m\omega x^{2}\geq2|px|\geq\hbar
\]
by $\omega/2$ we then get
\[
E=\frac{p^{2}}{2m}+\frac{1}{2}m\omega^{2}x^{2}\geq\frac{1}{2}\hbar\omega
\]
which is the correct lower bound for the quantum energy. The argument above
can also be reversed: since the lowest energy of an oscillator with frequency
$\omega$ and mass $m$ is $\frac{1}{2}\hbar\omega$, the minimal phase space
trajectory will be the ellipse%
\[
\frac{p^{2}}{m\hbar\omega}+\frac{x^{2}}{(\hbar/m\omega)}=1\text{;}%
\]
that ellipse encloses an area equal to $\frac{1}{2}h$, which is a topological,
or geometrical, version of the uncertainty principle. Everything in the
discussion above immediately extends to the $n$-dimensional oscillator with
phase space coordinates $(x_{1},...,x_{n};p_{1},...,p_{n})$ by using each of
the uncertainty relations $\Delta p_{j}\Delta x_{j}\geq\frac{1}{2}\hbar$, and
one not only recovers the correct ground energy level, but one also finds that
conversely, the projection of the motion on any plane of conjugate variables
$x_{j},p_{j}$ will always enclose a surface having an area at least equal to
$\frac{1}{2}h$.

These heuristic observations leading to \emph{exact} results suggest that
there might be a precise relation between the uncertainty principle and the
ground energy level in more general cases. The aim of this paper is to show
that one can in fact use with benefit recent advances in symplectic topology
(Gromov's \cite{gromov} surprising ``non-squeezing theorem'' and the
``symplectic capacities'' of Ekeland and Hofer \cite{EH}), to both extend and
make rigorous the considerations above, not only for quadratic Hamiltonians,
but also for Liouville-integrable Hamiltonian systems.

It is not taking too great risks to conjecture that these new symplectic
methods --which were still unknown to mathematicians only two decades ago--
will play in the future a fundamental role in physics, both classical and
quantum. In \cite{select} and \cite{paselect} we already discussed EBK
quantization from the perspective of symplectic capacities; our argument
however relied on an \textit{ad hoc} physical assumption: part of the present
paper makes these results mathematically rigorous.

This paper is to a great extent self-contained; symplectic non-squeezing
results are, for the time being, not widely known by physicists (and perhaps
not even fully appreciated outside specialized mathematical circles): we have
therefore devoted Section \ref{secone} of this paper to an (elementary) review
of these recent advances; we prove, in passing, a linear version of Gromov's
theorem (Proposition \ref{un}) using a (probably) new approach. EBK
quantization of Lagrangian manifolds is also discussed in some detail, and a
precise definition of the Maslov index is given. For a technical mathematical
study of non-squeezing and general Lagrangian manifold we refer to our
previous paper \cite{physlett}. We also note that Dragoman \cite{Dragoman} has
used the related notion of quantum blob we have introduced in \cite{ICP} to
propose an axiomatic construction of quantum mechanics in phase space.
\medskip

\noindent\textsc{Notation}. We will use the following notation in this paper.
The phase space $\mathbb{R}_{z}^{2n}\equiv\mathbb{R}_{x}^{n}\times
\mathbb{R}_{p}^{n}$ is equipped with the standard symplectic form
\[
\sigma(z,z^{\prime})=p\cdot x^{\prime}-p^{\prime}\cdot x
\]
($z=(x,p)$, $z^{\prime}=(x^{\prime},p^{\prime})$); in differential notation:
\[
dp\wedge dx=dp_{1}\wedge dx_{1}+\cdot\cdot\cdot+dp_{n}\wedge dx_{n}%
\]
where $x=(x_{1},...,x_{n})$, $p=(p_{1},...,p_{n})$. We will call each pair
$(x_{j},p_{j})$ a pair of \emph{conjugate coordinates}. The symplectic group
of $(\mathbb{R}_{z}^{2n},\sigma)$ is denoted by $Sp(n)$: it is the group of
all linear automorphisms of $\mathbb{R}_{z}^{2n}$ such that $\sigma
(Sz,Sz^{\prime})=\sigma(z,z^{\prime})$ for all $(z,z^{\prime})$. We will call
a diffeomorphism $f:\mathbb{D}\subset\mathbb{R}_{z}^{2n}\longrightarrow
\mathbb{R}_{z}^{2n}$ a \emph{symplectomorphism} if the Jacobian matrix $Df(z)$
is in $Sp(n)$ for every $z\in\mathbb{D}$. The Lagrangian Grassmannian of
$(\mathbb{R}_{z}^{2n},\sigma)$ is denoted by $\Lambda(n)$; it is the manifold
of all Lagrangian planes in $(\mathbb{R}_{z}^{2n},\sigma)$, i.e. of the
$n$-dimensional subspaces of $\mathbb{R}_{z}^{2n}$ on which $\sigma$ is
identically zero. A Lagrangian manifold is a manifold whose tangent spaces are
Lagrangian planes.

A solution $t\longmapsto z(t)=(x(t),p(t))$ of the Hamilton equations%
\[
\dot{x}(t)=\partial_{p}H(z(t))\text{ \ , \ }\dot{p}(t)=-\partial_{x}H(z(t))
\]
for $H\in$ $C^{\infty}(\mathbb{R}_{z}^{2n},\mathbb{R)}$ will be called
indifferently ``solution curve'' or ``motion''.

We will denote by $B(\bar{z}^{\prime},R)$ the Euclidean phase space ball
$|z-\bar{z}^{\prime}|\leq R$ and by $S_{j}^{1}(\bar{z},r)$ the circle in the
conjugate plane $x_{j},p_{j}$ plane with radius $r$ and centered at $\bar{z}$.
The phase space cylinder $S_{j}^{1}(\bar{z},r)\times\mathbb{R}_{z}^{2n}$ based
on the $x_{j},p_{j}$ plane is denoted by $Z_{j}(\bar{z},r)$. We will write
$B(0,R)=B(R)$, $S_{j}^{1}(0,r)=S_{j}^{1}(0,r)$, and $Z_{j}(0,r)=Z_{j}(r)$.

\section{Symplectic Non-Squeezing Theorems\label{secone}}

The determinant of a symplectic matrix is equal to one; it follows that
symplectomorphisms are volume preserving: this is essentially the message of
Liouville's theorem on conservation of phase space volume by Hamiltonian
flows; it is however not a characteristic of Hamiltonian systems: Liouville's
theorem holds for the flow of any divergence-free vector field. What really
singles out symplectomorphisms among all volume-preserving diffeomorphisms is
the following ``non-squeezing property'' proved by Gromov \cite{gromov} in
1985. One way of expressing Gromov's theorem is to say that for every
symplectomorphism $f$ defined in a neighbourhood of $B(\bar{z}^{\prime},R)$,
the area of the orthogonal projection of $f(B(\bar{z}^{\prime},R))$ on any of
the conjugate planes $x_{j},p_{j}$ will have an area which is at least equal
to that of the projection of $B(\bar{z}^{\prime},R)$ itself on that plane,
that is $\pi R^{2}$. It follows from this statement that there exists no
symplectomorphism $f$ such that $f(B(\bar{z}^{\prime},R))\subset Z_{j}(\bar
{z},r)$ if $R>r$. (That there exist such symplectomorphisms if $R\leq r$ is
obvious: translations in $\mathbb{R}_{z}^{2n}$ are trivially symplectic).

All known proofs of Gromov's theorem rely on rather complicated mathematical
methods (e.g. the theory of pseudo-holomorphic curves). Here is however a
proof in the affine case (an affine symplectomorphism is the compose of an
element of $Sp(n)$ and of a translation in $\mathbb{R}_{z}^{2n}$). Since phase
space translations trivially satisfy Gromov's theorem, we may, without loss of
generality, reduce the proof to the case $\bar{z}=\bar{z}^{\prime}=0$. We are
in fact going to show that for every $S\in Sp(n)$ the area of the orthogonal
projection of $S(B(R))$ (where $B(R)=B(0,R)$) on any of the conjugate planes
$x_{j},p_{j}$ is $\geq\pi R^{2}$.

\begin{proposition}
\label{un}Let $S\in Sp(n)$. (1) The area of the intersection of $S(B(R))$ with
any of the conjugate planes $x_{j},p_{j}$ is equal to $\pi R^{2}$; (2) The
area of the orthogonal projection of $S(B(R))$ on any of the conjugate planes
$x_{j},p_{j}$ is at least $\pi R^{2}$.
\end{proposition}

\begin{proof}
The second statement follows from the first since the orthogonal projection of
$S(B(R))$ on the $x_{j},p_{j}$ plane contains the intersection of $S(B(R))$
with that plane. Let us prove (1). The area of the plane surface
$\Gamma=S(B(R))\cap\mathbb{R}_{x_{j},p_{j}}^{2}$ is
\[
A(\Gamma)=\oint_{\gamma}p_{j}dx_{j}=\oint_{\gamma}pdx
\]
where $\gamma$ is the (positively) oriented boundary of $\Gamma$ and $pdx$ the
Liouville form $p_{1}dx_{1}+\cdot\cdot\cdot+p_{n}dx_{n}$. Since $S$ is linear
the set $\Gamma^{\prime}=S^{-1}(\Gamma)$ is a surface lying in a plane passing
through the origin, and its boundary $\gamma^{\prime}=S^{-1}(\gamma)$ is hence
a big circle of the sphere $|z|=R$. Using Stoke's theorem together with the
fact that the symplectic form $dp\wedge dx$ is preserved by $S$ we have
\[
A(\Gamma)=\int_{\Gamma}dp\wedge dx=\int_{S^{-1}(\Gamma)}dp^{\prime}\wedge
dx^{\prime}%
\]
so it is sufficient to show that
\begin{equation}
\int_{S^{-1}(\Gamma)}dp^{\prime}\wedge dx^{\prime}=\pi R^{2}\text{.}
\label{aprou}%
\end{equation}
A new application of Stoke's formula yields
\[
\int_{S^{-1}(\Gamma)}dp^{\prime}\wedge dx^{\prime}=\oint_{\gamma^{\prime}%
}p^{\prime}dx^{\prime}\text{;}%
\]
parametrizing $\gamma^{\prime}$ by
\begin{gather*}
x_{j}^{\prime}(t)=x_{j}^{\prime}\cos t+p_{j}^{\prime}\sin t\\
p_{j}^{\prime}(t)=-x_{j}^{\prime}\sin t+p_{j}^{\prime}\cos t\\
\sum_{j=1}^{n}(x_{j}^{\prime2}+p_{j}^{\prime2})=R^{2}%
\end{gather*}
($1\leq j\leq n$\ ,\ $0\leq t\leq1$) we have
\[
\oint_{\gamma^{\prime}}p^{\prime}dx^{\prime}=\pi\sum_{j=1}^{n}(x_{j}^{\prime
2}+p_{j}^{\prime2})=\pi R^{2}%
\]
proving (\ref{aprou}) and hence the proposition.
\end{proof}

\begin{remark}
It would certainly be interesting to extend the proof above along the same
lines to the case of arbitrary symplectomorphisms. This might perhaps be
achieved by exploiting the fact that the image of a conjugate plane by any
symplectomorphism is a two-dimensional symplectic manifold.
\end{remark}

Gromov's theorem is equivalent to the existence of \emph{symplectic
capacities}. A symplectic capacity (for short: \textit{capacity}) on
$\mathbb{R}_{z}^{2n}$ is the assignment $c:\Omega\longmapsto c(\Omega)$ to
every subset $\Omega$ of $\mathbb{R}_{z}^{2n}$ of a number $\geq0$, or
$+\infty$, satisfying the following axioms:

\begin{itemize}
\item $\Omega\subset\Omega^{\prime}\Longrightarrow c(\Omega)\leq
c(\Omega^{\prime})$ for all $\Omega,\Omega^{\prime}\subset$ $\mathbb{R}%
_{z}^{2n}$;

\item $c(\lambda\Omega)=\lambda^{2}c(\Omega)$ for all $\lambda\in\mathbb{R}$;

\item $c(f(\Omega))=c(\Omega)$ for every symplectomorphism $f:\mathbb{R}%
_{z}^{2n}\longrightarrow\mathbb{R}_{z}^{2n}$;

\item $c(B(R))=\pi R^{2}=c(Z_{j}(R))$.
\end{itemize}

The first and fourth axioms obviously imply the following very useful
property:
\begin{equation}
B(R)\subset\Omega\subset Z_{j}(R)\Longrightarrow c(\Omega)=\pi R^{2}
\label{inclus}%
\end{equation}
for every symplectic capacity $c$.

In the case $n=1$ (the phase plane), the usual notion of area is a symplectic
capacity (for measurable sets); in higher dimensions volume is however never a
capacity (the second axiom would be violated); it seems that there is no
useful relation between volumes and capacities for $n>1$: property
(\ref{inclus}) shows that sets with very different sizes and volumes (even
infinite) can have the same capacity.

The existence of symplectic capacities is actually equivalent to Gromov's
theorem. This can be seen by introducing the lower and upper ``Gromov
capacities'' $c_{\mathrm{G}}$ and $c^{\mathrm{G}}$. They are defined as
follows: $c_{\mathrm{G}}(\Omega)=\pi R^{2}$ where $R$ is the supremum of the
radii of all balls that can be sent in $\Omega$ using symplectomorphisms;
$c^{\mathrm{G}}(\Omega)$ is the infimum of the radii of all cylinders
$Z_{j}(R)$ into which $\Omega$ can be sent using symplectomorphisms. The first
three axioms above are trivially satisfied by $c_{\mathrm{G}}$ and
$c^{\mathrm{G}}$; the fourth axiom is a restatement of Gromov's theorem. One
moreover easily checks that $c_{\mathrm{G}}$ and $c^{\mathrm{G}}$ are lower
and upper bounds for all capacities: we have
\begin{equation}
c_{\mathrm{G}}(\Omega)\leq c(\Omega)\leq c^{\mathrm{G}}(\Omega) \label{enca}%
\end{equation}
for all $\Omega\subset\mathbb{R}^{2n}$ and every symplectic capacity $c$.

Although there is at this time no general formula allowing the calculation of
the capacities of arbitrary sets, there are some partial results. Here are two
that will be used in this paper.

\textit{Ellipsoids}. Let $Q$ be a positive definite quadratic form on
$\mathbb{R}_{z}^{2n}$. There exists a linear symplectomorphism $S$ and a
unique $n$-tuplet $(R_{1},...,R_{n})$ of numbers $>0$ (the ``symplectic
spectrum of $Q$'') such that
\[
Q(Sz)=\sum_{j=1}^{n}\frac{1}{R_{j}^{2}}(x_{j}^{2}+p_{j}^{2})
\]
All the capacities of the ellipsoid $Q(z)\leq1$ are equal and are given by the
formula
\begin{equation}
c(B(R_{1},...,R_{n}))=\pi\inf_{1\leq j\leq n}R_{j}^{2} \label{ellipsoid}%
\end{equation}
(see e.g. Hofer--Zehnder \cite{HZ} for a proof).

\textit{Solid Lagrangian tori}. A solid Lagrangian torus is a product%
\[
\mathbb{D}^{n}(R_{1},...,R_{n})=D_{1}^{2}(R_{1})\times\cdot\cdot\cdot\times
D_{n}^{2}(R_{n})
\]
where $D_{j}^{2}(R_{j})$ is the disk $x_{j}^{2}+p_{j}^{2}\leq R_{j}^{2}$ lying
in the $x_{j},p_{j}$ plane. To calculate $c(\mathbb{D}^{n}(R_{1},...,R_{n}))$
we proceed as follows: let $B(R_{1},...,R_{n})$ be the ellipsoid defined by%
\[
\sum_{j=1}^{n}\frac{1}{R_{j}^{2}}(x_{j}^{2}+p_{j}^{2})\leq1\text{.}%
\]
We have inclusions%
\[
B(R_{1},...,R_{n})\subset\mathbb{D}^{n}(R_{1},...,R_{n})\subset Z_{j}(R_{j})
\]
for every $j=1,2,...,n$ and hence%
\[
c(B(R_{1},...,R_{n}))\leq c(\mathbb{D}^{n}(R_{1},...,R_{n}))\leq c(Z_{j}%
(R_{j}))\text{.}%
\]
In view of (\ref{ellipsoid}) and the equality $c(Z_{j}(R_{j}))=\pi R_{j}^{2}$
we get, using the third and fourth axioms for symplectic capacities,%
\[
\pi\inf_{1\leq j\leq n}R_{j}^{2}\leq c(\mathbb{D}^{n}(R_{1},...,R_{n}))\leq\pi
R_{j}^{2}%
\]
for $j=1,2,...,n$; choosing in particular $j$ such that $R_{j}^{2}=\inf_{1\leq
j\leq n}R_{j}^{2}$ it follows that all the capacities of the solid torus are
equal and are given by%
\begin{equation}
c(\mathbb{D}^{n}(R_{1},...,R_{n}))=\pi\inf_{1\leq j\leq n}R_{j}^{2}\text{.}
\label{soto}%
\end{equation}

\section{Quadratic Hamiltonians\label{quad}}

Let $H$ be a positive definite quadratic form in the position and momentum
variables, that is%
\[
H(z)=\frac{1}{2}Rz\cdot z=\frac{1}{2}z^{T}Rz
\]
where $R$ is a real symmetric $2n\times2n$ matrix with $>0$ eigenvalues (it is
the Hessian matrix of $H$). Let $S\in Sp(n)$ be such that
\[
H(Sz)=\sum_{j=1}^{n}\frac{1}{R_{j}}(p_{j}^{2}+x_{j}^{2})
\]
where $R_{1},...,R_{n}>0$; setting $\omega_{j}=\sqrt{2/R_{j}}$, the compose
$H\circ S$ can be written in the familiar form
\[
H(Sz)=\sum_{j=1}^{n}\frac{\omega_{j}}{2}(p_{j}^{2}+x_{j}^{2})\text{;}%
\]
notice that the frequencies $\omega_{j}$ are uniquely determined by $H$ and
are thus independent of the choice of $S$. Solving Hamilton's equations for
$H\circ S$ with initial datum $(x,p)$ at time $t=0$ yields
\begin{align*}
x_{j}(t)  &  =x_{j}\cos\omega_{j}t+p_{j}\sin\omega_{j}t\\
p_{j}(t)  &  =-x_{j}\sin\omega_{j}t+p_{j}\cos\omega_{j}t
\end{align*}
$=\allowbreak\left(  x_{j}t=x_{j}\cos\omega_{j}t+p_{j}\sin\omega_{j}%
t,p_{j}t=-x_{j}\sin\omega_{j}t+p_{j}\cos\omega_{j}t\right)  $ for $1\leq j\leq
n$ and hence the motion winds around a torus%
\begin{equation}
\mathbb{T}^{n}=S_{1}^{1}(R_{1})\times\cdot\cdot\cdot\times S_{n}^{1}%
(R_{n})\text{ \ , \ }R_{j}=\sqrt{x_{j}^{2}+p_{j}^{2}}\text{.} \label{tor1}%
\end{equation}
We know from standard quantum mechanics that the exact quantized energy levels
of $H$ are given by the formula%
\[
E_{N_{1},...,N_{n}}=\sum_{j=1}^{n}(N_{j}+\tfrac{1}{2})\hbar\omega_{j}%
\]
where the $N_{j}$ are integers $\geq0$; in particular the ground energy level
is
\begin{equation}
E_{0}=\sum_{j=1}^{n}\tfrac{1}{2}\hbar\omega_{j}\text{.} \label{gr}%
\end{equation}
In quantum mechanics this property is usually restated by saying that ``a sum
of harmonic oscillators is in its ground energy level if and only if\ each of
its components is''. Let us discuss formula (\ref{gr}) from a semiclassical
perspective. The fact that each individual oscillator with Hamiltonian%
\[
H_{j}(x_{j},p_{j})=\frac{\omega_{j}}{2}(p_{j}^{2}+x_{j}^{2})
\]
has ground energy $\tfrac{1}{2}\hbar\omega_{j}$ means that the corresponding
semiclassical motion takes place on the circle
\[
\frac{\omega_{j}}{2}(p_{j}^{2}+x_{j}^{2})=\tfrac{1}{2}\hbar\omega_{j}%
\]
with radius $R_{j}=\sqrt{\hbar}$. It follows that the motion determined by the
complete Hamiltonian $H=H_{1}+\cdot\cdot\cdot+H_{n}$ is carried by the torus%
\[
\mathbb{T}^{n}(\sqrt{\hbar})=S_{1}^{1}(\sqrt{\hbar})\times\cdot\cdot
\cdot\times S_{n}^{1}(\sqrt{\hbar})\text{.}%
\]
It follows from formula (\ref{soto}) of Section \ref{secone} that every
capacity of the solid torus
\[
\mathbb{D}^{n}(\sqrt{\hbar})=D_{1}^{2}(\sqrt{\hbar})\times\cdot\cdot
\cdot\times D_{n}^{2}(\sqrt{\hbar})
\]
is equal to
\begin{equation}
c(\mathbb{D}^{n}(\sqrt{\hbar}))=\pi(\sqrt{\hbar})^{2}=\tfrac{1}{2}h\text{.}
\label{cdn}%
\end{equation}

\begin{remark}
\label{remo}In the light of Gromov's theorem (\ref{cdn}) can be viewed as a
topological form of the uncertainty principle: choosing
$c=c^{_{\text{\textrm{G}}}}$, (\ref{cdn}) shows that $\mathbb{D}^{n}%
(\sqrt{\hbar})$ cannot be squeezed inside a cylinder $Z_{j}(R)$ with radius
$R<\sqrt{\hbar}$using symplectomorphisms.
\end{remark}

Suppose now, conversely, that the motion takes place on a torus $\mathbb{T}%
^{n}(R_{1},...,R_{n})$ and that the capacity of the corresponding solid torus
is
\begin{equation}
c(\mathbb{D}^{n}(R_{1},...,R_{n}))=\tfrac{1}{2}h\text{.} \label{cd1}%
\end{equation}
Using again formula (\ref{soto}) we get
\[
c(\mathbb{D}^{n}(R_{1},...,R_{n}))=\pi R^{2}=\tfrac{1}{2}h
\]
where $R=\inf_{1\leq j\leq n}R_{j}$. It follows that we have $R_{j}^{2}%
\geq\hbar$ for $j=1,2,...,n$ ; the energy of the motion being given by%
\[
E=\sum_{j=1}^{n}\frac{\omega_{j}}{2}(p_{j}^{2}+x_{j}^{2})=\sum_{j=1}^{n}%
\frac{\omega_{j}}{2}R_{j}^{2}%
\]
it follows that the assumption (\ref{cd1}) implies that $E\geq E_{0}$ where
$E_{0}$ is the ground energy level\ (\ref{gr}), and thus implies the correct
lower bound for the energy.

Summarizing:

\begin{itemize}
\item A necessary condition for the motion of a positive definite quadratic
Hamiltonian to be quantized is that it lies on a torus $\mathbb{T}^{n}$ such
that the corresponding solid torus $\mathbb{D}^{n}$ has symplectic capacity
$c(\mathbb{D}^{n})$ at least equal to $\frac{1}{2}h$, that is half the quantum
of action.

\item The condition that the motion is carried by a torus $\mathbb{T}^{n}$
such that $c(\mathbb{D}^{n})$ is not sufficient to conclude that this motion
is quantized; its energy is however bounded from below by the ground energy level.
\end{itemize}

Let us extend this discussion to a class of more general Hamiltonian systems.

\section{Liouville-Integrable Hamiltonian Systems}

Let $H$ be a Hamiltonian function on $\mathbb{R}_{z}^{2n}$; we assume that
there exists a symplectomorphism $f:(x,p)\longmapsto(\phi,I)$ of
$\mathbb{R}_{z}^{2n}$ (not necessarily globally defined) such that $K=H\circ
f^{-1}$ only depends on the variables $I=(I_{1},...,I_{n})$:
\[
H(x,p)=K(I)\text{.}%
\]
The Hamilton equations for $K$ (and hence for $H$) are immediately solved, and
one finds that%
\begin{equation}
\phi_{j}(t)=\omega_{j}(I(0))t+\phi_{j}(0)\text{ \ , \ }I_{j}(t)=I_{j}(0)\text{
\ for \ }1\leq j\leq n\text{;} \label{aa}%
\end{equation}
the frequencies $\omega_{j}$ are the derivatives of $K$:
\begin{equation}
\omega_{j}(I)=\partial_{I_{j}}K(I)\text{ \ \ , \ \ }1\leq j\leq n\text{.}
\label{om}%
\end{equation}
Such a situation typically occurs when the Hamiltonian system associated to
$H$ is Liouville integrable, that is when: (1) there exist $n$ independent
constants of the motion $F_{1}=H,F_{2},...,F_{n}$ so that the set
\[
\mathbb{V}^{n}=\{z:F_{1}(z)=f_{1},...,F_{n}(z)=f_{n}\}\text{;}%
\]
is a $n$-dimensional manifold for almost all values $f_{1},...,f_{n}$ of
$F_{1},...,F_{n}$; (2) these constants of the motion are in involution:%
\[
\{F_{i},F_{j}\}=\partial_{x}F_{i}\cdot\partial_{p}F_{j}-\partial_{x}F_{j}%
\cdot\partial_{p}F_{i}=0
\]
hence the manifold $\mathbb{V}^{n}$ is Lagrangian; each motion takes place on
such a manifold. In fact, the manifolds $\mathbb{V}^{n}$ can be parametrized
the $n$-parameter $I=(I_{1},...,I_{n})$ consisting of the ``action
variables'', at least in some open subset $U\subset\mathbb{R}^{n}$. If the
manifolds $\mathbb{V}^{n}$ are compact and connected there exists a
symplectomorphism $f:(x,p)\longmapsto(\phi,I)$, defined in a neighbourhood of
$\mathbb{V}^{n}$, such that $f(\mathbb{V}^{n})$ is the torus
\[
\mathbb{T}^{n}(R_{1},...,R_{n})=S_{1}^{1}(R_{1})\times\cdot\cdot\cdot\times
S_{n}^{1}(R_{n})
\]
(the variables $\phi=(\phi_{1},...,\phi_{n})$ are here cyclic).

The passage to semiclassical mechanics consists in imposing selection rules on
the Lagrangian manifolds $\mathbb{V}^{n}$; these rules are the EBK
(Einstein--Brillouin--Keller) quantum conditions:%
\begin{align}
&  \frac{1}{2\pi\hbar}\oint_{\gamma}pdx-\frac{1}{4}m(\gamma)\text{ \ is an
integer}\label{EBK1}\\
&  \text{for all one-cycles }\gamma\text{ on }\mathbb{V}^{n}\nonumber
\end{align}
(see e.g. Arnol'd, Leray \cite{Leray}, Maslov \cite{Maslov}, Maslov--Fedoriuk
\cite{MF}; the conditions (\ref{EBK1}) are sometimes also called the
Bohr--Sommerfeld--Maslov conditions in the literature). The integer
$m(\gamma)$ appearing in (\ref{EBK1}) is the Maslov index of $\gamma$; its
vocation is to ``count'' the number of caustics of $\mathbb{V}^{n}$ traversed
by $\gamma$ (a caustic of $\mathbb{V}^{n}$ is a point of $\mathbb{V}^{n}$
which does not have any neighbourhood diffeomorphic to an open subset of the
position space $\mathbb{R}_{x}^{n}$). The Maslov index is calculated as
follows (Arnol'd \cite{arnold}, Leray \cite{Leray}, Souriau \cite{Souriau2}).
Parametrize $\gamma$ by $t\in\lbrack0,1]$ and set $\ell(t)=T_{\gamma
(t)}\mathbb{V}^{n}$ (the tangent plane to $\mathbb{V}^{n}$ at $\gamma(t)$. The
mapping $t\longmapsto\ell(t)$, $0\leq t\leq1$, is a loop $\gamma_{\Lambda}$ in
the Lagrangian Grassmannian $\Lambda(n)$. Identifying $\Lambda(n)$ with the
manifold $W(n)$ of all symmetric unitary matrices of order $n$ (see Souriau
\cite{Souriau2}; also Guillemin--Sternberg \cite{GS1}), the loop
$\gamma_{\Lambda}$ is identified with a loop $\gamma_{W}:$ $t\longmapsto
w(t)$, $0\leq t\leq1$, in $W(n)$. The Maslov index of $\gamma$ is by
definition the integer%
\[
m(\gamma)=\frac{1}{2\pi i}\int_{0}^{1}\frac{d(\det w(t))}{\det w(t)}\text{.}%
\]
One shows that $m(\gamma)$ only depends on the homotopy class of $\gamma$ in
$\mathbb{V}^{n}$. An important property of the Maslov index is the following:
\begin{equation}
\mathbb{V}^{n}\text{ oriented }\Longrightarrow m(\gamma)\text{ is even}
\label{vorien}%
\end{equation}
(Souriau \cite{Souriau3}).

The semiclassical values of the energy are obtained from the EBK condition as
follows: let $I=(I_{1},...,I_{n})$ be the action variables corresponding to
the basic one-cycles $\gamma^{1},...,\gamma^{n}$ on $\mathbb{V}^{n}$. These
are defined as follows: let $\bar{\gamma}^{1},...,\bar{\gamma}^{n}$ be the
loops in $\mathbb{T}^{n}(R_{1},...,R_{n})$ defined, for $0\leq t\leq2\pi$, by
\begin{gather*}
\bar{\gamma}^{1}(t)=R_{1}(\cos t,0,...,0;\sin t,0,...,0)\\
\bar{\gamma}^{2}(t)=R_{2}(0,\cos t,...,0;0,\sin t,...,0)\\
\cdot\cdot\cdot\cdot\cdot\cdot\cdot\cdot\cdot\cdot\cdot\cdot\cdot\cdot
\cdot\cdot\cdot\cdot\cdot\cdot\cdot\cdot\cdot\cdot\\
\bar{\gamma}^{n}(t)=R_{n}(0,...,0,\cos t;0,...,0,\sin t)\text{.}%
\end{gather*}
The basic one-cycles $\gamma^{1},...,\gamma^{1}$ of $\mathbb{V}^{n}$ are then
just
\[
\gamma^{1}=f^{-1}(\bar{\gamma}^{1}),...,\gamma^{n}=f^{-1}(\bar{\gamma}%
^{n})\text{.}%
\]
The action variables being given by
\[
I_{j}=\frac{1}{2\pi}\oint_{\bar{\gamma}^{j}}Id\phi=\frac{1}{2\pi}\oint
_{\gamma^{j}}pdx\text{ \ , \ }1\leq j\leq n
\]
the EBK quantization conditions (\ref{EBK1}) imply that we must have%
\begin{equation}
I_{j}=(N_{j}+\tfrac{1}{4}m(\gamma^{j}))\hbar\text{ \ \ for \ \ }1\leq j\leq n
\label{quantac}%
\end{equation}
each $N_{j}$ being an integer $\geq0$. Writing $H(x,p)=K(I)$ the semiclassical
energy levels are then given by the formula%
\begin{equation}
E_{N_{1},...,N_{n}}=K((N_{1}+\tfrac{1}{4}m(\gamma^{1}))\hbar,...,(N_{n}%
+\tfrac{1}{4}m(\gamma^{1}))\hbar) \label{eleve}%
\end{equation}
where $N_{1},...,N_{n}$ range over all \emph{non-negative} integers; they
correspond to the physical \textquotedblleft quantum states\textquotedblright%
\ labeled by the sequence $(N_{1},...,N_{n})$.

(We do not discuss here the ambiguity that might arise in the calculation of
the energy because of the non-uniqueness of the angle action coordinates; that
ambiguity actually disappears if one requires that the system under
consideration is non-degenerate, that is $\partial^{2}K(I)\neq0$.)

Let us state and prove a result which generalizes to the Liouville integrable
case the discussion of quadratic Hamiltonians we did in the last Section.

\begin{theorem}
\label{deux}Assume that the Lagrangian manifold $\mathbb{V}^{n}$ is compact
and connected. (1) If $\mathbb{V}^{n}$ satisfies the EBK condition
(\ref{EBK1}), then for every symplectic capacity $c$ we have
\begin{equation}
c(\mathbb{\bar{V}}^{n})\geq\tfrac{1}{2}h \label{cv}%
\end{equation}
where $\mathbb{\bar{V}}^{n}\ $is defined by $f(\mathbb{\bar{V}}^{n}%
)=\mathbb{D}^{n}$ ($f$ the mapping $(x,p)\longmapsto(\phi,I)$) and
$\mathbb{D}^{n}$ is the ``solid torus'' corresponding to $\mathbb{T}^{n}$. (2)
If conversely $\mathbb{\bar{V}}^{n}$ satisfies (\ref{cv}), and the frequencies
$\omega_{j}$ are everywhere $>0$ then the energy $E$ of the motion carried by
$\mathbb{V}^{n}$ is such that%
\begin{equation}
E\geq E_{0}=K(\tfrac{1}{2}\hbar,...,\tfrac{1}{2}\hbar)\text{;} \label{est}%
\end{equation}
In view of (\ref{b}) we have $m(\gamma^{j})\geq2$ for every basic one-cycle
$\gamma^{j}$ on $\mathbb{V}^{n}$ and hence%
\begin{equation}
E_{N_{1},...,N_{n}}\geq E_{0}=K(\tfrac{1}{2}\hbar,...,\tfrac{1}{2}%
\hbar)\text{;} \label{esta}%
\end{equation}
the number $E_{0}$ is a lower bound for the quantized energy levels
$E_{N_{1},...,N_{n}}$ given by (\ref{eleve}).
\end{theorem}

\begin{proof}
(1) Since capacities are symplectic invariants, we may assume without
restricting the generality of the argument that $\mathbb{V}^{n}$ is the torus
$\mathbb{T}^{n}=\mathbb{T}^{n}(R_{1},...,R_{n})$ itself. Since
\[
I_{j}=\frac{1}{2\pi}\oint_{\gamma^{j}}pdx=\frac{1}{2\pi}\oint_{\bar{\gamma
}^{j}}I_{j}d\phi_{j}=\frac{1}{2}R_{j}^{2}%
\]
the quantization conditions (\ref{quantac}) are equivalent to the conditions%
\[
R_{j}^{2}=(2N_{j}+\tfrac{1}{2}m(\gamma^{j}))\hbar\text{.}%
\]
As a manifold $\mathbb{V}^{n}$ (and hence $\mathbb{T}^{n}$) has dimension $n$;
we must thus have $R_{j}>0$ for every $j$, and this implies that $m(\gamma
^{j})>0$ for every basic one-cycle $\gamma^{j}$. It follows that%
\begin{equation}
\inf_{1\leq j\leq n}R_{j}^{2}\geq\tfrac{1}{2}\inf_{1\leq j\leq n}m(\gamma
^{j}))\hbar>0\text{.} \label{a}%
\end{equation}
We next observe that the torus $\mathbb{T}^{n}=\mathbb{T}^{n}(R_{1}%
,...,R_{n})$ is an oriented manifold (because it is a product of circles,
which are oriented manifolds). It follows that $\mathbb{V}^{n}=f^{-1}%
(\mathbb{T}^{n})$ is also oriented (symplectomorphisms are orientation
preserving). Souriau's theorem (\ref{vorien}) thus implies that the Maslov
index $m(\gamma^{j})$ of every basic one-cycle on $\mathbb{V}^{n}$ is even,
and hence%
\begin{equation}
\inf_{1\leq j\leq n}m(\gamma^{j})\geq2\text{.} \label{b}%
\end{equation}
It follows from the inequalities (\ref{a}) and (\ref{b}) that we have%
\[
c(\mathbb{D}^{n}(R_{1},...,R_{n}))=\pi\inf_{1\leq j\leq n}R_{j}^{2}\geq
\tfrac{1}{2}h
\]
as was to be proven. (2) Assume that conversely
\[
c(\mathbb{\bar{V}}^{n})=c(\mathbb{D}^{n})\geq\tfrac{1}{2}h\text{.}%
\]
The motion thus takes place on a torus $\mathbb{T}^{n}=\mathbb{T}^{n}%
(R_{1},...,R_{n})$ such
\[
\pi\inf_{1\leq j\leq n}R_{j}^{2}\geq\tfrac{1}{2}h
\]
and we thus have%
\[
I_{j}=\frac{1}{2\pi}\oint_{\bar{\gamma}^{j}}I_{j}d\phi_{j}=\frac{1}{2}%
R_{j}^{2}\geq\frac{1}{2}\hbar\text{.}%
\]
The assumption $\omega_{j}(I)=\partial_{I_{j}}K(I)>0$ implies that $K$ is an
increasing function of the variables $I=(I_{1},...,I_{n})$ and we thus have%
\[
E=K(I_{1},...,I_{n})\geq K(\tfrac{1}{2}\hbar,...,\tfrac{1}{2}\hbar)\text{.}%
\]
In view of (\ref{b}) we have $m(\gamma^{j})\geq2$ for every basic one-cycle
$\gamma^{j}$ on $\mathbb{V}^{n}$ and hence%
\[
E_{N_{1},...,N_{n}}\geq K(\tfrac{1}{2}\hbar,...,\tfrac{1}{2}\hbar)
\]
which ends the proof of the Theorem.
\end{proof}

\begin{remark}
If we impose the EBK conditions on the tori $\mathbb{T}^{n}$ themselves, and
not on the Lagrangian manifolds $\mathbb{V}^{n}$ then the number $E_{0}$ in
(\ref{est}) effectively coincides with the ground energy level; but in the
general case we have $E_{0}<E_{N_{1},...,N_{n}}$. This is due to the fact that
the Maslov index is not a symplectic invariant and that $\mathbb{V}^{n}$ can
have more caustics than $\mathbb{T}^{n}$ (see the discussion above following
the definition of the Maslov index).
\end{remark}

From Theorem \ref{deux} we easily deduce the following form of the uncertainty principle:

\begin{corollary}
\label{trois}Let $\mathbb{V}^{n}$ be a compact and connected Lagrangian
manifold associated to a\ Liouville-integrable Hamiltonian system. If
$\mathbb{V}^{n}$ satisfies the EBK condition (\ref{EBK1}) then the following
property holds: Let $\Lambda_{j}$ be a connected subset in the $x_{j},p_{j}$
plane bounded by a simple curve\textit{ }$\lambda_{j}$. If $\Lambda_{j}$
\textit{contains the projection of }$\mathbb{V}^{n}$ then we have\textit{ }%
\begin{equation}
\operatorname{Area}(\Lambda_{j}))\geq\frac{1}{2}h\text{\textit{.}}
\label{area}%
\end{equation}
\textit{ }
\end{corollary}

\begin{proof}
Recall from Theorem \ref{deux} that if $\mathbb{V}^{n}$ is quantized then
$c(\mathbb{\bar{V}}^{n})\geq\tfrac{1}{2}h$. Let us first assume that
$\lambda_{j}$ is a circle $S_{j}^{1}(R)$. Then $\mathbb{\bar{V}}^{n}\subset
Z_{j}(R)$ and hence
\[
\tfrac{1}{2}h\leq c(\mathbb{\bar{V}}^{n})\leq c(Z_{j}(R))=\pi R^{2}%
\]
proving the claim in that case. If $\lambda_{j}$ is not a circle, choose an
area-preserving diffeomorphism $f$ of the $x_{j},p_{j}$\textit{\ }plane taking
$\Lambda_{j}$ into a circle $S_{j}^{1}(R)$. The phase space transformation $F$
taking $(x_{j},p_{j})$ into $f(x_{j},p_{j})$ and leaving all other coordinates
unchanged is symplectic, and the projection of $F(\mathbb{\bar{V}}^{n})$ lies
inside $S_{j}^{1}(R)$. We have
\[
c(\mathbb{\bar{V}}^{n})=c(F(\mathbb{\bar{V}}^{n}))\leq c(Z_{j}%
(R))=\operatorname{Area}(\Lambda_{j})
\]
hence\ again $\operatorname{Area}(\Lambda_{j})\geq\frac{1}{2}h$.
\end{proof}

\begin{remark}
This result can also be deduced from Gromov's theorem: if $c(\mathbb{\bar{V}%
}^{n})\geq\tfrac{1}{2}h$ for every symplectic capacity $c$ then, in
particular, $c_{\mathrm{G}}(\mathbb{V}^{n})\geq\tfrac{1}{2}h$ ($c_{\mathrm{G}%
}$ the Gromov capacity defined in Section \ref{secone}). The largest ball
$B(R)$ that can be squeezed inside $\mathbb{V}^{n}$ by a symplectomorphism $f$
has therefore radius $\sqrt{\hbar}$, and the area of the orthogonal projection
of $f(B(\sqrt{\hbar}))$ on any $x_{j},p_{j}$ plane is at least $\frac{1}{2}h$.
\end{remark}

\begin{remark}
The inequality (\ref{area}) is symplectically invariant in the sense that it
remains true if we replace $\mathbb{\bar{V}}^{n}$ by $g(\mathbb{\bar{V}}^{n}%
)$, $g$ any symplectomorphism: this follows from the symplectic invariance of
symplectic capacities.
\end{remark}

\section{Discussion and Conclusion}

We have been able to relate the Heisenberg uncertainty principle to the
existence of a non-zero ground energy level for integrable systems with
compact Lagrangian tori. The results we have obtained are however not sharp,
in the sense that we have not been able to recover the exact ground energy
from the minimum uncertainty, but only a lower bound for that energy. A
possible way to refine and generalize these results would perhaps be to use
the powerful tool of ``Hofer displacement energy'' (see Hofer--Zehnder
\cite{HZ}, Polterovich \cite{Poltero}).

In \cite{select,ICP} we have shown that a classical uncertainty principle,
formally identical with the quantum uncertainty principle, can be derived for
all linear Hamiltonian systems; it would perhaps be interesting to extend
these results to more general Hamiltonians (integrable or not), and to study
the \emph{classical} implications of this principle from the point of view of
the methods outlined in this paper: perhaps the existence of these classical
uncertainty relations could be used with some profit in the study of
non-integrable (chaotic) Hamiltonian systems.

Gromov's theorem, and its implications, shows that Hamiltonian mechanics is
``aerial'' in nature; symplectic capacities are symplectic invariants that
have the physical dimension of an area, that is of \textit{action}. They
certainly deserve to be further studied within the contexts of both classical
and quantum mechanics. A possible application of the notion of symplectic
capacity might be a global characterization of adiabatic invariance (and of
the method of ``adiabatic switching'' in semiclassical mechanics). One might
envisage that in multi-dimensional Hamiltonian systems the best candidate for
adiabatic conservation is not the action of periodic orbits, but rather the
capacity of some sets (for instance that of Lagrangian solid tori in the
integrable case, or that of the set bounded by the energy shell in the ergodic
case). We hope to come back to these important topics in forthcoming work.

\begin{acknowledgement}
This work has been supported by a generous grant 2005/51766-7 of the research
agency FAPESP of the State of Sa\~{o} Paulo (Brazil). I would like to take the
opportunity to express my warmest thanks to Professor Paolo Piccione (IME,
University of S\~{a}o Paulo) for his kind invitation and for having provided
the author with a very congenial working environment. I would also like take
the opportunity to thank Basil Hiley (Birkbeck College, London) for
stimulating conversations on foundational issues in quantum mechanics.
\end{acknowledgement}


\begin{thebibliography}{99}                                                                                               %


\bibitem {arnold}\textsc{Arnold, V I} \textit{A characteristic class entering
in quantization conditions}, Funkt. Anal. i. Priloz. \textbf{1}(1) (1967)
1--14 (in Russian); Funct. Anal. Appl. \textbf{1} (1967) 1--14 (English translation)

\bibitem {Arnold}\textsc{Arnold, V I} \textit{Mathematical Methods of
Classical Mechanics}, 2d edition, Graduate Texts in Mathematics,
Springer--Verlag (1978)

\bibitem {Dragoman}\textsc{D Dragoman} Phase Space Formulation of Quantum
Mechanics. Insight into the Measurement Problem. \textit{Phys. Scr}.
\textbf{72} (2005) 290-296

\bibitem {EH}\textsc{Ekeland, I and Hofer, H} \textit{Symplectic topology and
Hamiltonian dynamics}, I and II, Math. Zeit. \textbf{200} (1990) 355--378 and
\textbf{203} (1990) 553--67

\bibitem {select}\textsc{de Gosson, M} \textit{The symplectic camel and phase
space quantization}, J. Phys. A: Math. Gen. \textbf{34} (2001) 1085--89

\bibitem {paselect}\textsc{de Gosson, M} \textit{The `symplectic camel
principle' and semiclassical mechanics}, J. Phys. A: Math. Gen \textbf{35}
(2002) 6825--6851

\bibitem {ICP}\textsc{de Gosson, M} \textit{The Principles of Newtonian and
Quantum Mechanics; with a foreword by B. Hiley, }Imperial College Press (2001)

\bibitem {physlett}\textsc{de Gosson, M} \textit{Phase Space Quantization and
the Uncertainty Principle.} Phys. Lett. A, \textbf{317}:365--369, 2003.

\bibitem {gromov}\textsc{Gromov, M} \textit{Pseudoholomorphic curves in
symplectic manifolds}, Invent. Math. \textbf{82} (1985) 307--47

\bibitem {GS1}\textsc{Guillemin, V, and Sternberg, } \textit{Geometric
Asymptotics}, Math. Surveys Monographs \textbf{14}, Amer. Math. Soc.,
Providence RI (1978)

\bibitem {HZ}\textsc{Hofer, H, and Zehnder, E} \textit{Symplectic Invariants
and Hamiltonian Dynamics}, Birkh\"{a}user Advanced texts (Basler
Lehrb\"{u}cher, Birkh\"{a}user Verlag) (1994)

\bibitem {Leray}\textsc{Leray, J} \textit{Lagrangian Analysis and Quantum
Mechanics},\textit{\ a mathematical structure related to asymptotic expansions
and the Maslov index}, MIT Press, Cambridge, Mass. (1981)

\bibitem {Maslov}\textsc{Maslov, V P} \textit{Th\'{e}orie des Perturbations et
M\'{e}thodes Asymptotiques}, Dunod, Paris (1972); [original Russian edition: 1965]

\bibitem {MF}\textsc{Maslov, V P, and Fedoriuk, M V} \textit{Semi-Classical
Approximations in Quantum Mechanics}, Reidel, Boston (1981)

\bibitem {Poltero}\textsc{Polterovich, E} \textit{The Geometry of the Group of
Symplectic Diffeomorphisms}, Lectures in Mathematics, ETH\ Z\"{u}rich
Birkh\"{a}user (2001)

\bibitem {Souriau2}\textsc{Souriau, J M} \textit{Construction explicite de
l'indice de Maslov}, Group Theoretical Methods in Physics, Lecture Notes in
Physics, \textbf{50}, Springer-Verlag (1975) 17--148

\bibitem {Souriau3}\textsc{Souriau, J M} \textit{Indice de Maslov des
vari\'{e}t\'{e}s lagrangiennes orientables}, C. R. Acad. Sci., Paris,
S\'{e}rie A, \textbf{276} (1973) 1025--26
\end{thebibliography}
\end{document}